\documentclass[a4paper]{svmult}
\usepackage{times}
\usepackage[dvips]{graphicx,epsfig}
\usepackage{amsmath,amsfonts,amssymb}
\usepackage{cite,url}

\begin{document}



\mainmatter

\title*{The Missing Three-Nucleon Forces: Where Are They?}

\author{R.~Machleidt\thanks{Talk presented at the 28th Intern.\ Workshop on
Nuclear Theory, Rila Mountains, Bulgaria, June 22-27, 2009.}}

\titlerunning{The Missing Three-Nucleon Forces: Where Are They?}
\authorrunning{R.~Machleidt}

\toctitle{The Missing Three-Nucleon Forces: Where Are They?}
\tocauthor{R.~Machleidt}

\institute{Department of Physics, University of Idaho, Moscow, Idaho 83844, USA}

\maketitle

\begin{abstract}
In recent years, there has been substantial progress in the derivation of nuclear forces
from chiral effective field theory. Accurate two-nucleon forces (2NF) have been constructed
up to next-to-next-to-next-to-leading order (N$^3$LO)
of chiral perturbation theory
and applied in microscopic nuclear structure calculations with a good degree of success. 
However, chiral three-nucleon forces (3NF) have been used only at N$^2$LO, improving
some miscroscopic predictions, but leaving also several issues, 
like the ``$A_y$ puzzle''
of nucleon-deuteron scattering, unresolved. Thus, the 3NF at N$^3$LO is needed for
essentially two reasons:
For consistency with the 2NF, and
to (hopefully) improve some critical predictions of nuclear structure  and reactions.
However, there are indications that the 3NF at N$^3$LO (in the so-called
$\Delta$-less version of the theory) is rather weak and may not solve
any of the outstanding problems. 
If this suspicion is confirmed, we have to go beyond, which may be similar
to opening Pandora's Box. In this talk, I will discuss the various possible scenarios 
and how to deal with them.
\end{abstract}

\section{Introduction}

The problem of a proper derivation of nuclear forces is as old as nuclear
phsyics itself, namely, almost 80 years~\cite{Mac89,Mac06}.
The modern view is that,
since the nuclear force is a manifestation of strong interactions, any serious derivation 
has to start from quantum chromodynamics (QCD). 
However, the well-known problem with QCD is that it is non-perturbative 
in the low-energy regime characteristic for nuclear physics.
For many years this fact was perceived as the great obstacle for a derivation
of nuclear forces from QCD---impossible to overcome except by lattice QCD.
The effective field theory (EFT) concept has shown the way out of this dilemma.
One has to realize that the scenario of low-energy QCD is characterized by pions
and nucleons interacting via a force governed by spontaneously broken approximate
chiral symmetry.
This chiral EFT allows for a systematic low-momentum
expansion known as chiral perturbation theory (ChPT)~\cite{Wei79}. 
Contributions are analyzed in terms of powers of small momenta
over the large scale: $(Q/\Lambda_\chi)^\nu$, where $Q$ is generic for a
momentum (nucleon three-momentum or pion four-momentum) or pion mass and
$\Lambda_\chi \approx 1$ GeV is the chiral symmetry breaking scale.
The early applications of ChPT focused on systems like $\pi\pi$~\cite{GL84}
and $\pi N$~\cite{GSS88}, where the Goldstone-boson character of the pion
guarantees that the expansion converges.

The past 15 years have also seen great progress in applying ChPT to nuclear forces
\cite{Wei90,Wei92,ORK94,Kol94,Kol99,KBW97,KGW98,EGM98,BK02,EM02,EM02a,EM03,ME05,EGM05,Mac07}.
As a result, nucleon-nucleon ($NN$) potentials of high precision have been constructed which
are based upon ChPT carried to next-to-next-to-next-to-leading order 
(N$^3$LO)~\cite{EM03,EGM05,Mac07}. Thus, the ground work for the derivation of nuclear forces
from chiral EFT is laid and the attention now turns to more detailed conceptual
questions as well as the construction of higher order corrections.
To be more specific, the most crucial open issues in the field of chiral nuclear forces
are
\begin{itemize}
\item
the renormalization of chiral nuclear potentials and
\item
subleading chiral few-nucleon forces.
\end{itemize}
This talk is devoted to the latter issue.
The renormalization problem is discussed in length elsewhere in the 
literature~\cite{Ent08}.

I will first provide a general overview of how nuclear forces emerge from
chiral EFT (Section 2) and then discuss in more detail the specific issue of
sub-leading few-nucleon forces (Section 3). Section 4 contains a summary
and a prospect for the future.

\section{Nuclear forces from chiral EFT: Overview}

\subsection{Chiral perturbation theory and power counting
\label{sec_chpt}}

Effective Langrangians have infinitely
many terms, and an unlimited number of Feynman graphs can be calculated
from them. Therefore, 
we need a scheme that makes the theory manageable and calculabel.
This scheme
which tells us how to distinguish between large
(important) and small (unimportant) contributions
is chiral perturbation theory (ChPT), and
determining the power $\nu$ of the expansion
has become known as power counting.

Nuclear potentials are defined as sets of irreducible
graphs up to a given order.
The power $\nu$ of a few-nucleon diagram involving $A$ nucleons
is given by:
\begin{equation} 
\nu = -2 +2A - 2C + 2L 
+ \sum_i \Delta_i \, ,  
\label{eq_nu} 
\end{equation}
with
\begin{equation}
\Delta_i  \equiv   d_i + \frac{n_i}{2} - 2  \, ,
\label{eq_Deltai}
\end{equation}
where $C$ denotes the number of separately connected pieces and
$L$ the number of loops in the diagram;
$d_i$ is the number of derivatives or pion-mass insertions and $n_i$ the number of nucleon fields 
(nucleon legs) involved in vertex $i$; the sum runs over all vertices contained
in the diagram under consideration.
Note that $\Delta_i \geq 0$
for all interactions allowed by chiral symmetry.
For an irreducible $NN$ diagram (``two-nucleon force'', $A=2$, $C=1$),
Eq.~(\ref{eq_nu}) collapses to
\begin{equation} 
\nu =  2L + \sum_i \Delta_i \, .  
\label{eq_nunn} 
\end{equation}

The power formula 
Eq.~(\ref{eq_nu}) 
allows to predict
the leading orders of multi-nucleon forces.
Consider a $m$-nucleon irreducibly connected diagram
($m$-nucleon force) in an A-nucleon system ($m\leq A$).
The number of separately connected pieces is
$C=A-m+1$. Inserting this into
Eq.~(\ref{eq_nu}) together with $L=0$ and 
$\sum_i \Delta_i=0$ yields
$\nu=2m-4$. Thus, two-nucleon forces ($m=2$) start 
at $\nu=0$, three-nucleon forces ($m=3$) at
$\nu=2$ (but they happen to cancel at that order),
and four-nucleon forces at $\nu=4$ (they don't cancel).
Thus, ChPT provides a straightforward explanation for
the empirically known fact that 2NF $\gg$ 3NF $\gg$ 4NF
\ldots.

In summary, the chief point of the ChPT expansion is
that,
at a given order $\nu$, there exists only a finite number
of graphs. This is what makes the theory calculable.
The expression $(Q/\Lambda_\chi)^{\nu+1}$ provides a rough estimate
of the relative size of the contributions left out and, thus,
of the accuracy at order $\nu$.
In this sense, the theory can be calculated to any
desired accuracy and has
predictive power.

\subsection{The hierarchy of nuclear forces}
\begin{figure}[t]
\vspace*{-0.5cm}
\scalebox{0.55}{\includegraphics{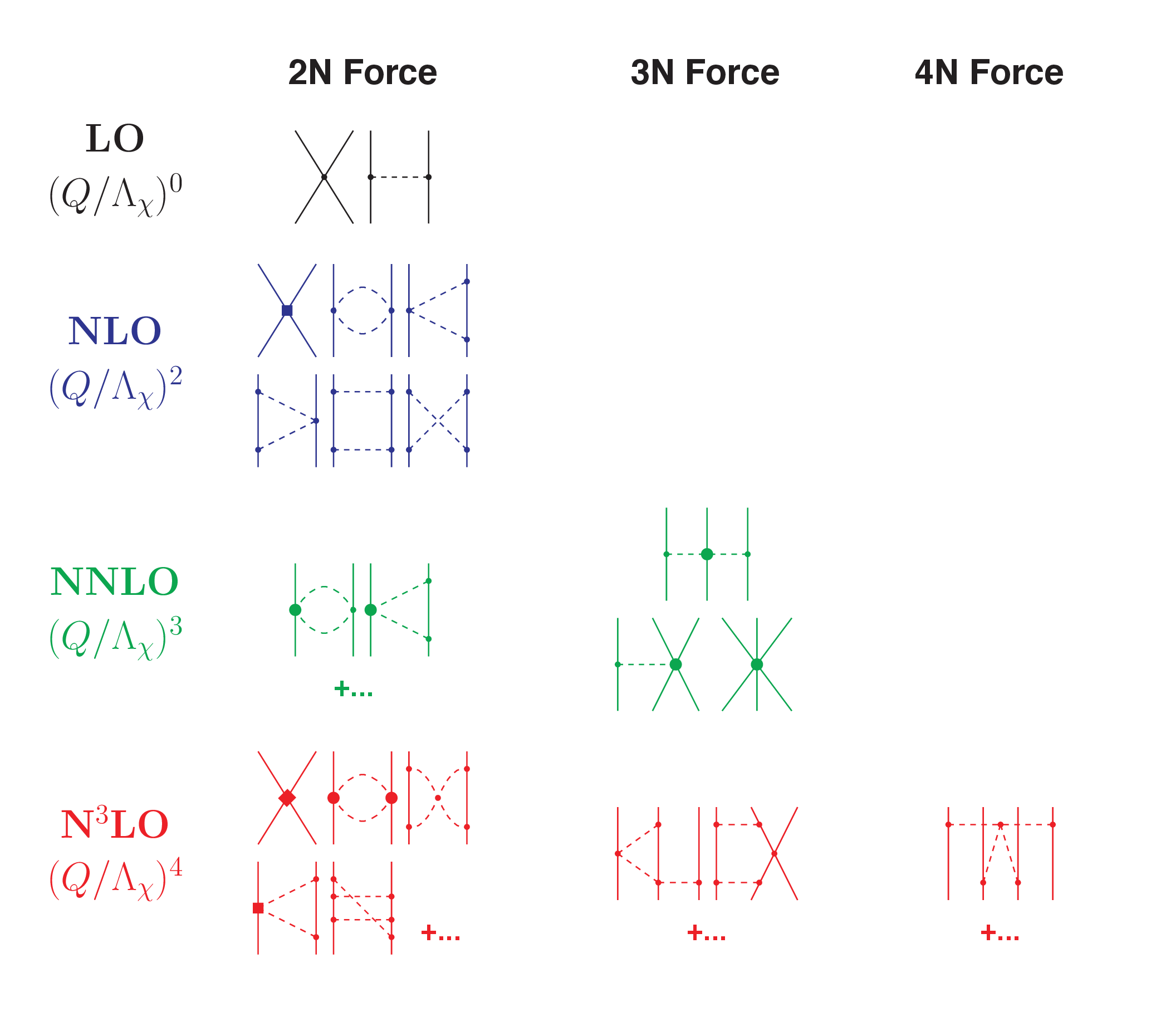}}
\vspace*{-0.50cm}
\caption{Hierarchy of nuclear forces in ChPT. Solid lines
represent nucleons and dashed lines pions. 
Small dots, large solid dots, solid squares, and solid diamonds
denote vertices of index $\Delta= \, $ 0, 1, 2, and 4, respectively. 
Further explanations are
given in the text.}
\label{fig_hi}
\end{figure}

Chiral perturbation theory and power counting
imply that nuclear forces emerge as a hierarchy
controlled by the power $\nu$, Fig.~\ref{fig_hi}.

In lowest order, better known as leading order (LO, $\nu = 0$), 
the $NN$ amplitude
is made up by two momentum-independent contact terms
($\sim Q^0$), 
represented by the 
four-nucleon-leg graph
with a small-dot vertex shown in the first row of 
Fig.~\ref{fig_hi},
and
static one-pion exchange (1PE), second
diagram in the first row of the figure.
This is, of course, a rather crude approximation
to the two-nucleon force, but accounts already for some
important features.
The 1PE provides the tensor force,
necessary to describe the deuteron, and it explains
$NN$ scattering in peripheral partial waves of very high
orbital angular momentum. At this order, the two contacts 
which contribute only in $S$-waves provide
the ``intermediate-range'' attraction which, indeed,
is a rather rudimentary description of reality.

In the next order,
$\nu=1$, all contributions vanish due to parity
and time-reversal invariance.

Therefore, the next-to-leading order (NLO) is $\nu=2$.
Two-pion exchange (2PE) occurs for the first time
(``leading order 2PE'') and, thus, the creation of
a more realistic
description of the intermediate-range attraction
is starting here. 
Since the loop involved in each pion-diagram implies
already $\nu=2$ [cf.\ Eq.~(\ref{eq_nunn})],
the vertices must have $\Delta_i = 0$.
Therefore, at this order, only the lowest order
$\pi NN$ and $\pi \pi NN$ vertices are allowed which
is why the leading order 2PE is rather weak.
Furthermore, there are 
seven contact terms of 
${\cal O}(Q^2)$, 
shown by
the four-nucleon-leg graph with a solid square,
 which contribute
in $S$ and $P$ waves. The operator structure of these
contacts include a spin-orbit term besides central,
spin-spin, and tensor terms. Thus, essentially all spin-isospin
structures necessary to describe the two-nucleon
force phenomenologically have been generated at this order.
The main deficiency at this stage of development 
is an insufficient intermediate-range attraction.

This problem is finally fixed at order three 
($\nu=3$), next-to-next-to-leading order (NNLO).
The 2PE involves now the $\Delta_i=1$
$\pi\pi NN$ seagull vertices (proportional to
the $c_i$ LECs) denoted by a large solid dot
in Fig.~\ref{fig_hi}.
These vertices represent correlated 2PE
as well as intermediate $\Delta(1232)$-isobar contributions.
It is well-known from the meson phenomenology of 
nuclear forces~\cite{Mac89,MHE87}
that these two contributions are crucial
for a realistic and quantitative 2PE model.
Consequently, the 2PE now assumes a realistic size
and describes the intermediate-range attraction of the
nuclear force about right. Moreover, first relativistic 
corrections come into play at this order.
There are no new contacts.

The reason why we talk of a hierarchy of nuclear forces is that 
two- and many-nucleon forces are created on an equal footing
and emerge in increasing number as we go to higher and higher orders.
At NNLO, the first set of
nonvanishing three-nucleon forces (3NF) occur~\cite{Kol94,Epe02b},
cf.\ column `3N Force' of
Fig.~\ref{fig_hi}. 
In fact, at the previous order, NLO,
irreducible 3N graphs appear already, however,
it has been shown by Weinberg~\cite{Wei92} and 
others~\cite{Kol94,YG86,CF86} that these diagrams all cancel.
Since nonvanishing 3NF contributions happen first
at order 
$(Q/\Lambda_\chi)^3$, 
they are very weak as compared to the 2NF which starts at
$(Q/\Lambda_\chi)^0$.

More 2PE is produced at $\nu =4$, next-to-next-to-next-to-leading
order (N$^3$LO), of which we show only a few symbolic diagrams in 
Fig.~\ref{fig_hi}. 
Two-loop 2PE
graphs show up for the first time and so does
three-pion exchange (3PE) which necessarily involves
two loops.
3PE was found to be negligible at this order~\cite{Kai00}.
Most importantly, 15 new contact terms $\sim Q^4$
arise and are represented 
by the four-nucleon-leg graph with a solid diamond.
They include a quadratic spin-orbit term and
contribute up to $D$-waves.
Mainly due to the increased number of contact terms,
a quantitative description of the
two-nucleon interaction up to about 300 MeV
lab.\ energy is possible
at N$^3$LO~\cite{EM03}. 
Besides further 3NF,
four-nucleon forces (4NF) start
at this order. Because the leading order 4NF 
comes into existence one
order higher than the leading 3NF, 4NFs are weaker
than 3NFs.

Since 2003, a very quantitative chiral $NN$ potential (at N$^3$LO)~\cite{EM03}
exists which has been applied successfully in many nuclear 
structure 
calculations~\cite{Cor02,FOS04,Nav07,Hag08,Bog05}.
Therefore, the chiral two-nucleon force appears to be
in good shape (except for the renormalization issue discussed
elsewhere~\cite{Ent08}). However, there are still open questions in the few-nucleon-force
sector as we will explain now in more detail.

\section{Few-nucleon forces \label{sec_manyNF}}

Nuclear three-body forces in ChPT were initially discussed
by Weinberg~\cite{Wei92}.
The 3NF at NNLO, was derived by van Kolck~\cite{Kol94}
and applied, for the first time, in nucleon-deuteron
scattering by Epelbaum {\it et al.}~\cite{Epe02b}.
The leading 4NF (at N$^3$LO) was recently constructed by
Epelbaum~\cite{Epe06} and found to contribute in the
order of 0.1 MeV to the $^4$He binding energy
(total $^4$He binding energy: 28.3 MeV)
in a preliminary calculation~\cite{Roz06}, confirming the traditional
assumption that 4NF are essentially negligible.
{\bf Therefore, the focus is on 3NF.}

For a 3NF, we have $A=3$ and $C=1$ and, thus, Eq.~(\ref{eq_nu})
implies for 3NF
\begin{equation}
\nu = 2 + 2L + 
\sum_i \Delta_i \,.
\label{eq_nu3nf}
\end{equation}
We will use this equation to analyze 3NF contributions
order by order.
The lowest possible power is obviously $\nu=2$ (NLO), which
is obtained for no loops ($L=0$) and 
only leading vertices
($\sum_i \Delta_i = 0$). 
This 3NF happens to vanish~\cite{Wei92}.
The first non-vanishing 3NF occurs at NNLO.

\begin{figure}[t]\centering
\vspace*{-5.50cm}
\hspace*{-1.0cm}
\scalebox{0.65}{\includegraphics{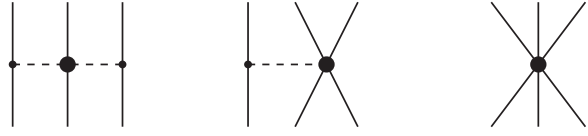}}
\vspace*{-12.00cm}
\caption{The three-nucleon force at NNLO.
From left to right: 2PE, 1PE, and contact diagrams.
Notation as in Fig.~\ref{fig_hi}.}
\label{fig_3nf_nnlo}
\end{figure}

\subsection{The 3NF at NNLO}
The power $\nu=3$ (NNLO) is obtained when
there are no loops ($L=0$) and 
$\sum_i \Delta_i = 1$, i.e., 
$\Delta_i=1$ for one vertex 
while $\Delta_i=0$ for all other vertices.
There are three topologies which fulfill this condition,
known as the two-pion exchange (2PE), 1PE,
and contact graphs
(Fig.~\ref{fig_3nf_nnlo}).
In this figure, vertices represented by a small dot carry
$\Delta_i=0$ while large solid dots have $\Delta_i=1$.

The 3NF at NNLO (Fig.~\ref{fig_3nf_nnlo}) 
has been evaluated (without the $1/M_N$ corrections)~\cite{Kol94,Epe02b}
and applied in
calculations of few-nucleon reactions~\cite{Epe02b,KE07,Viv08},
structure of light- and medium-mass 
nuclei~\cite{Cor02,FOS04,Nav07,Hag08},
and nuclear matter~\cite{Bog05}
with a fair deal of success.
However, the famous `$A_y$ puzzle' of nucleon-deuteron scattering
is not solved~\cite{Epe02b,KE07}, and the even bigger problem with the
analyzing power in $p$-$^3$He scattering~\cite{Fis06,DF07} will certainly not be fixed
at this order.
Furthermore, the spectra of light nuclei leave room for improvement~\cite{Nav07}.

We note that there are further 3NF contributions at NNLO, namely, the
$1/M_N$ corrections of the NLO 3NF diagrams.
Part of these corrections 
have been calculated by Coon and Friar in 1986~\cite{CF86}.
These contributions are believed to be very small.

In summary, because of various unresolved problems in microscopic nuclear structure, 
the 3NF beyond NNLO is very much in need. In fact, it is no exaggeration to state that
the 3NF at sub-leading orders
is presently one of the most important outstanding issues in the chiral EFT
approach to nuclear forces.

\begin{figure}[t]\centering
\vspace*{-6.0cm}
\scalebox{0.70}{\includegraphics{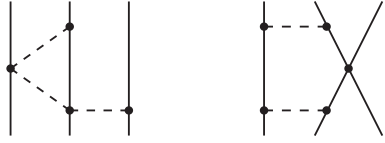}}
\vspace*{-12.8cm}
\caption{The 3NF at N$^3$LO:
Two examples of one-loop graphs.
Notation as in Fig.~\ref{fig_hi}.}
\label{fig_3nf_n3lo}
\end{figure}

\subsection{The 3NF at N$^3$LO}

According to Eq.~(\ref{eq_nu3nf}),
the value $\nu=4$, which corresponds to N$^3$LO, is obtained
for the following classes of diagrams.

\subsubsection{3NF loop diagrams at N$^3$LO.}
For this group of graphs, we have
$L=1$ and, therefore, all $\Delta_i$ have to be zero
to ensure $\nu=4$. 
Thus, these one-loop 3NF diagrams can include
only leading order vertices, the parameters of which
are fixed from $\pi N$ and $NN$ analysis.
We show two samples
of this very large class of diagrams
in Fig.~\ref{fig_3nf_n3lo}. 
One sub-group of these diagrams (``$2\pi$ exchange graphs'')
has been calculated by Ishikawa and Robilotta~\cite{IR07},
and two other topologies ($2\pi$-$1\pi$ and ring diagrams)
have been evaluated by the Bonn-J\"ulich group~\cite{Ber08}.
The remaining topologies, which involve a leading order four-nucleon
contact term (e.g., second diagram of Fig.~\ref{fig_3nf_n3lo}),
are under construction by the Bonn-J\"ulich group.
The N$^3$LO $2\pi$-exchange 3NF has been applied in the calculation
of nucleon-deuteron observables in Ref.~\cite{IR07} 
producing very small effects.

The smallness of the $2\pi$ loop 3NF at N$^3$LO is not unexpected.
It is consistent with
experience with corresponding 2NF diagrams: 
the NLO 2PE contribution to the $NN$ potential, which 
involves one loop and only leading vertices, is also relatively small.

By the same token, one may expect that also all the other N$^3$LO
3NF loop topologies will produce only small effects.

\subsubsection{3NF tree diagrams at N$^3$LO.}
The order $\nu=4$ is also obtained for the combination $L=0$ (no loops)
and $\sum_i \Delta_i = 2$.
Thus, either two vertices have to carry $\Delta_i=1$ or
one vertex has to be of the $\Delta_i=2$ kind,
while all other vertices are $\Delta_i=0$.
This is achieved if 
in the NNLO 3NF graphs of Fig.~\ref{fig_3nf_nnlo}
the power of one vertex is raised by one.
The latter happens if a relativistic
$1/M_N$ correction is applied.
A closer inspection reveals that all $1/M_N$ corrections of the
NNLO 3NF vanish and the first non-vanishing corrections
are proportional to $1/M_N^2$ and appear at N$^4$LO.
However, there are non-vanishing $1/M_N^2$ corrections of the NLO 3NF
and there are so-called drift corrections~\cite{Rob06} 
which contribute at N$^3$LO (some drift corrections are claimed to
contribute even at NLO~\cite{Rob06}). 
We do not expect these contributions to be sizable.
Moreover, there are contributions from the $\Delta_i =2$
Lagrangian~\cite{FMS98} proportional to the
low-energy constants $d_i$. As it turns out, these terms have
at least one time-derivative, which causes them to be
$Q/M_N$ suppressed and demoted to N$^4$LO.

Thus, besides some minor $1/M_N^2$ corrections, there are no tree
contributions to the 3NF at N$^3$LO.

\subsubsection{Summarizing the entire N$^3$LO 3NF contribution:}
For the reasons discussed, we anticipate that this 3NF is weak and will not solve
any of the outstanding problems. 
In view of this expectation, we have to look for
more sizable 3NF contributions elsewhere.

\subsection{The 3NF at N$^4$LO of the $\Delta$-less theory}
The obvious step to take is to proceed to the next order,
N$^4$LO or $\nu=5$, of the $\Delta$-less theory which is the one
we have silently assumed so far. (The $\Delta$-full theory will
be introduced and discussed below.)
Some of the tree diagrams that appear at this order were mentioned already:
the $1/M_N^2$ corrections of the NNLO 3NF and the trees with one $d_i$
vertex which are $1/M_N$ suppressed. Because of the suppression factors,
we do not expect sizable effects from these graphs.
Moreover, there are also tree diagrams with one vertex from the
$\Delta_i=3$ $\pi N$ Lagrangian~\cite{Fet00,FM00} proportional to the 
LECs $e_i$. Because of the high dimension of these vertices and assuming
reasonable convergence, we do not anticipate much from these trees either.

\begin{figure}[t]\centering
\vspace*{-1.5cm}
\hspace*{-1.5cm}
\scalebox{0.70}{\includegraphics{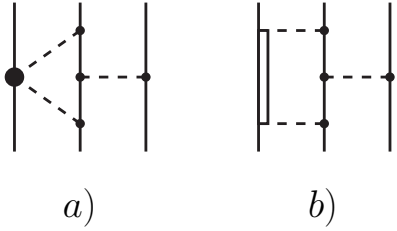}}
\vspace*{-15.00cm}
\caption{(a) One-loop 3NF at N$^4$LO of the $\Delta$-less theory.
(b) Corresponding diagram of the $\Delta$-full theory which contributes
at N$^3$LO.
Double lines represent $\Delta$ isobars; other
notation as in Fig.~\ref{fig_hi}.}
\label{fig_3nf_n4lo}
\end{figure}

However, we believe that the loop contributions that occur at this order are truly important.
They are obtained by replacing in the N$^3$LO loops (Fig.~\ref{fig_3nf_n3lo})
one vertex by a $\Delta_i=1$ vertex [with LEC $c_i$]. 
We show one symbolic example of this large group of diagrams
in Fig.~\ref{fig_3nf_n4lo}(a).
This 3NF is presumably large and, thus, what we are looking for.

The reasons, why these graphs are large, can be argued as follows.
Corresponding 2NF diagrams are the three-pion exchange (3PE)
contributions to the $NN$ interaction. In analogy to 
Figs.~\ref{fig_3nf_n3lo} and \ref{fig_3nf_n4lo}(a),
there are 3PE 2NF diagrams with only leading vertices and the ones with one (sub-leading) 
$c_i$ vertex (and the rest leading). These diagrams have been evaluated by Kaiser in 
Refs.~\cite{Kai00} and \cite{Kai01}, respectively. 
Kaiser finds that the 3PE contributions with one sub-leading vertex are about
an order magnitude larger then the leading ones.

\subsection{N$^3$LO 3NF contributions in the $\Delta$-full theory}

The above considerations indicate that the $\Delta$-less theory exhibits,
in some cases, a bad convergence pattern.
The reason for the unnaturally strong subleading contributions are the large
values of the $\Delta_i=1$ LECs, $c_i$. The large values can be explained
in terms of resonance saturation~\cite{BKM97}.
The $\Delta(1232)$-resonance contributes considerably to $c_3$ and $c_4$.
The explicit inclusion of the $\Delta$ takes strength out of these
LECs and moves this strength to a lower order, thus improving the
convergence~\cite{OK92,ORK94,KGW98,KEM07,EKM08}. 
Figure~\ref{fig_3nf_n4lo} illustrates this fact for the 3NF
under consideration: 
the diagram of the $\Delta$-less theory shown in (a)
is (largely) equivalent to diagram (b) which includes one $\Delta$ excitation.
Note, however,
that diagram (a) is N$^4$LO, while diagram (b) is N$^3$LO.
Moreover, there are further N$^3$LO one-loop diagrams with two and three $\Delta$ excitations,
which correspond to diagrams of order N$^5$LO and N$^6$LO, respectively, in the $\Delta$-less
theory. 

This consideration clearly shows that the inclusion of $\Delta$ degrees
of freedom in chiral EFT makes the calculation of sizable higher-order 3NF 
contributions much more efficient.

\section{Summary, Conclusions and Outlook}

The past 15 years have seen great progress in our understanding of nuclear forces
in terms of low-energy QCD. Key to this development was the realization that
low-energy QCD is equivalent to an effective field theory (EFT) which allows for 
a perturbative expansion that has become know as chiral perturbation theory (ChPT).
In this framework, two- and many-body forces emerge on an equal footing and the empirical fact
that nuclear many-body forces are substantially weaker than the two-nucleon force
is explained automatically.

In spite of the great progress and success of the past 15 years, there are still some
unresolved issues that will need our attention in the near future. One problem is the
proper renormalization of the chiral two- and many-nucleon potentials. This has not been the
subject of my talk, but a thorough discussion together with a comprehensive list of the
vast literature on the subject can be found in Ref.~\cite{Ent08}.

The other unfinished business are the few-nucleon forces beyond NNLO (``sub-leading
few-nucleon forces''). In this talk, we chose to elaborate on this topic, and
the bottom line can be summarized as follows:
\begin{itemize}
\item
The chiral 3NF at NNLO is insufficient. Additional {\it sizable}
3NF contributions are needed.
\item
The chiral 3NF at N$^3$LO (in the $\Delta$-less theory) 
most likely does {\it not} produce sizable contributions.
\item
Sizable contributions are expected from one-loop 3NF diagrams
at N$^4$LO of the $\Delta$-less or N$^3$LO of the $\Delta$-full
theory (Fig.~\ref{fig_3nf_n4lo}). {\it These 3NF contributions may
turn out to be the missing pieces in the 3NF puzzle and have the potential to solve
the outstanding problems in microscopic nuclear structure.}\footnote{Note that the Illinois
3NF~\cite{Pie01} includes two one-loop diagrams with one and two $\Delta(1232)$-isobars.
The deeper reason for this may be in arguments we are presenting.}
\end{itemize}

\section*{Acknowledgments}
It is my pleasure to acknowledge the instrumental role which D. R. Entem has played
in the field reviewed in this talk.
This work was supported in part by 
the US Department of Energy under
Grant No.\ DE-FG02-03ER41270.

\end{document}